\documentstyle[preprint,tighten,aps]{revtex}

\begin{document}
\draft
\preprint{IFUP-TH 2/99}
\title{
The Euclidean two-point correlation function of the topological charge density.
}
\author{Ettore Vicari}
\address{Dipartimento di Fisica dell'Universit\`a 
and I.N.F.N., I-56126 Pisa, Italy}

\date{\today}

\maketitle

\begin{abstract}
We study
the Euclidean two-point correlation function $G_q(x)$
of the topological charge density in QCD.
A general statement based on reflection positivity tells us that
$G_q(x)<0$ for $x\neq 0 $. On the other hand the topological
susceptibility $\chi_q=\int d^d x G_q(x)$ is a  positive quantity.
This indicates that $G_q(x)$ developes a positive contact term at $x=0$, 
that contributes to the
determination of the physical value of $\chi_q$.
We show explicitly these features of $G_q(x)$ in a solvable nontrivial continuum model,
the two-dimensional $CP^{N-1}$ model in the large-$N$ limit. 
A similar analysis is done on the lattice.


\end{abstract}

\newpage
%\newcommand{\N}{\hbox{{\rm I}\kern-.2em\hbox{\rm N}}}
% ========================= BODY =========================
%\narrowtext

\section{Introduction}
\label{introduction}

By the axial anomaly equation,
matrix elements and correlation functions 
involving the topological charge density operator
can be related to relevant quantities of the hadronic phenomenology.
The most important phenomenon related to the
topological properties of QCD is
 the explicit breaking of the $U(1)$ axial symmetry,
with the result that the theory contains neither a conserved
$U(1)$ axial quantum number, nor an extra Goldstone boson.

In  Euclidean space the topological charge density is 
\begin{equation}
q(x) = {g^2\over 32\pi^2} \epsilon_{\mu\nu\rho\sigma} 
{\rm Tr}\;F_{\mu\nu}(x) F_{\rho\sigma}(x).
\label{qx}
\end{equation}
The correlation function 
\begin{equation} 
G_q(x-y) \equiv  \langle q(x) q(y) \rangle  
\label{corrfunc}
\end{equation}
is important for understanding
the breaking of the $U(1)$ axial symmetry. 
According to Witten's argument~\cite{Witten} the breaking of the $U(1)$ 
axial symmetry should occur
at the lowest non-planar level, i.e. at the next-to-leading order
of the $1/N$ expansion, and require a nonzero large-$N$ limit of 
the topological susceptibility
\footnote{
The physically relevant topological susceptibility is correctly defined by Eq.~(\ref{chiqdef})
in terms of the Euclidean correlation function $G_q(x)$ (see e.g. Ref.~\cite{Meggiolaro}
and references therein).}
\begin{equation}
\chi_q = \int d^4 x \;G_q(x)
\label{chiqdef}
\end{equation}
in the pure $SU(N)$ gauge theory (see also the recent Ref.~\cite{Witten2}). 
An extension  of this idea relates $\chi_q$ of the large-$N$ 
pure gauge $SU(N)$ theory to the $\eta'$ mass~\cite{Veneziano}.
Many numerical studies, based on Monte Carlo simulations 
of lattice formulations of the pure gauge theory, have been devoted to the 
estimate of $\chi_q$ (see e.g. 
Refs.~\cite{D-F-R-V,L-gm,Tc,G-K-S-W,Sm-V,C-D-P,A-C-D-G-V,D-H-Z,A-D-D,D-G-S,N-V,Tn}).

Arguments based on reflection positivity~\cite{O-S-rf} tell us that $G_q(x)\leq 0$ 
for $|x|>0$ (this was already noted in Refs.~\cite{S-S,A-D-D-K}).
On the other hand $\chi_q \geq 0$ trivially from its definition.
These facts indicate that there is a positive 
contact term at $x=0$, that contributes
to determine the physical quantity $\chi_q$.
In order to investigate this issue, we consider the 
two-dimensional $CP^{N-1}$ model~\cite{D-D-L,Witten-2},
which is an interesting theoretical
laboratory for studying general topological properties.
We will calculate $G_q(x)$ in the large-$N$ limit,
providing an explicit analytical example where the main properties that should characterize 
$G_q(x)$ in QCD are realized.
As argued for QCD, $G_q(x)$ presents a short-distance singular behavior
characterized by a positive contact term at $x=0$ and a negative diverging approach
for $|x|\rightarrow 0$. 
Neverthless, the low-momentum behavior of $G_q(x)$
and in particular its moments
\begin{equation}
\chi_{q,j} \equiv \int d^dx \;(x^2)^j G_q(x)
\label{moments}
\end{equation}
(where $\chi_{q,0}\equiv \chi_q$)
turn out to be well defined and finite.

The paper is organized as follows. In Sec.~\ref{sec2} we discuss
some features of $G_q(x)$ in the QCD theory, 
that can be inferred using reflection positivity,
and other general arguments such as perturbation theory and renormalization group.
In Sec.~\ref{sec3}  
we compute the large-$N$ limit of $\langle q(x) q(y)\rangle$ in 
the two-dimensional $CP^{N-1}$ models. Its main features are discussed and compared with those of 
the correlation function $G_q(x)$ of QCD.
In Sec.~\ref{sec4} we study the continuum limit of 
correlation functions of lattice
discretizations of the topological charge density operator.
An explicit calculation is presented in the large-$N$ limit of a lattice formulation
of two-dimensional $CP^{N-1}$ models.

\section{ Reflection positivity}
\label{sec2}

In order that Euclidean correlation functions can be continued back to Minkowski space,
they have to obey a positivity condition: the so-called reflection 
positivity~\cite{O-S-rf,O-S}.
The general statement concerning reflection positivity is that
\begin{equation}
\langle ( \Theta F) F \rangle \geq 0,
\label{rfst}
\end{equation}
where $\Theta$ is the antilinear
reflection operator consisting in an  Euclidean time reflection
and a complex conjugation, and $F$ is an arbitrary 
gauge invariant function of the fields
having support only at positive Euclidean times
(see also Ref.~\cite{M-M}).
As a consequence of the intrinsic odd parity of $q(x)$ under reflection,
\begin{equation}
\Theta q(x)  = - q(\theta x) = - q(x_1,x_2,x_3,-x_4), 
\label{qtr}
\end{equation}
reflection positivity states that
\begin{equation}
G_q(x) \leq  0 \qquad {\rm for} \quad |x|>0.
\label{cqxd}
\end{equation}
This fact holds for any operator
that is intrisically odd with respect to reflection symmetry in the
Euclidean space.

The asymptotic large- and short-distance behaviors of $G_q(x)$
can be inferred  by general arguments.
At large distance $G_q(x)$ should decay exponentially 
(in the presence of fermions):
$G_q(x) \sim  e^{-m_{\eta'} r}$ apart from negative powers of $r\equiv |x|$.
Simple dimensional, perturbative and renormalization group arguments tell us that
for $r \rightarrow 0$
\begin{equation}
G_q(x) = {c\over r^{8}(\ln r )^2 }\left[ 1 + O\left({1\over \ln r}\right)\right], 
\label{r0beh}
\end{equation}
where $c$ is a negative constant. 
The logarithms can be related to a running coupling constant,
indeed in perturbation theory $G_q(x)$ is $O(g^2)$.
Since the topological susceptibility $\chi_q$ is positive 
($\chi_q=0$ in the presence of a massless fermion) and $G_q(x)<0$ for
$x\neq 0 $, $G_q(x)$ 
should develope a positive contact term at $x=0$, that
compensates the negative contribution of its integral for $x\neq 0$
and makes $\chi_q$ positive.
Inspite of this singular  short-distance behavior,
the low-momentum behavior and in particular the moments $\chi_{q,j}$ of $G_q(x)$,
are conjectured to be well defined and finite. 
In the next section  we
will see in a solvable model how these features can coexist.

In the lattice formulation of the theory one may define 
two versions of the reflection symmetry
that are equivalent in the continuum limit: site- and link-reflection simmetry.
We recall that reflection positivity is essential on the lattice for the 
existence of a self-adjoint Hamiltonian at finite lattice spacing
defined from the transfer matrix.
Site- and link-reflection symmetries are both 
satisfied by the Wilson lattice action of $SU(N)$ gauge theories~\cite{O-S},
and by Wilson fermions with Wilson parameter $r=1$~\cite{M-P}.
So Eq.~(\ref{rfst})
must hold also on the lattice for finite lattice spacing.
A lattice discretization  of the topological charge density operator
is for example~\cite{D-F-R-V}:
\begin{equation}
q_{L}(x)=- {1\over 2^4\times 32 \pi^2}
\sum^{\pm 4}_{\mu\nu\rho\sigma=\pm 1}
\epsilon_{\mu\nu\rho\sigma} {\rm Tr}
\left[ \Pi_{\mu\nu}\Pi_{\rho\sigma}\right],
\label{Q^L}
\end{equation}
where
$\Pi_{\mu\nu}(x) = U_\mu(x)U_\nu(x+\mu)U^\dagger_\mu(x+\nu)U^\dagger_\nu(x)$
is the plaquette operator and $U_\mu(x)$ is  the link variable, 
the sum is done over positive and
negative directions.
One may easily verify that $q_L(x)$ trasforms as $q(x)$, cf. Eq.(\ref{qtr}),
under site-reflection.
Again reflection positivity tells us that 
$G_q^L(x)\equiv \langle q_L(x) q_L(0) \rangle $ is negative for $x\neq 0$
(at least along the directions of the lattice and when 
there is no overlap among the link variables of the two operators). 
In the continuum limit the lattice correlation $G_q^L(x)$
should reproduce the  continuum correlation function $G_q(z)$.
This will be discussed later, in Sec.~\ref{sec4}.

\section{ The large-$\protect\bbox{N}$  limit of 
$\protect\bbox{\langle \lowercase{q}(\lowercase{x})
\lowercase{q}(\lowercase{y}) \rangle }$ 
in two-dimensional $\protect\bbox{CP^{N-1}}$ models.}
\label{sec3}

The arguments of the previous section can be also applied to the
two-dimensional $CP^{N-1}$ models as well. 
In the following we will consider their large-$N$ limit and will calculate the leading
order of the two-point correlation function of the corresponding topological charge density
(i.e. $O(1/N)\;$). We will show explicitly that
$G_q(x)$ develops a  singular behavior at the origin
consistently with the reflection positivity requirement 
 $G_q(x)<0$ for $x\neq 0$, and the positivity of the
corresponding topological susceptibility,
i.e. of its space integral.
Nevertheless, the low-momentum behavior of 
the $G_q(x)$ turns out to be  well
defined without the need of special subtractions.
This provides an explicit example where the conjectured main features of the 
two-point function $G_q(x)$ of QCD are verified.

$CP^{N-1}$ models are defined by the action~\cite{D-D-L,Witten-2}
\begin{equation}
S= {N\over 2g} \int d^2x\,\overline{D_\mu z}\, D_\mu z ,
\end{equation}
where $z$ is a $N$-component complex scalar field subject to the constraint $\bar{z}z=1$,
and the covariant derivative $D_\mu =\partial_\mu +iA_\mu$ is defined 
in terms of the composite gauge field
$A_\mu=i\bar{z}\partial_\mu z$.
Such models present a $U(1)$ gauge invariance related to the local
transformations:
$z(x)\rightarrow e^{i\alpha(x)}z(x)$ and
$A_\mu(x)\rightarrow A_\mu(x) - \partial_\mu \alpha(x)$.

Two-dimensional $CP^{N-1}$ models are interesting  
because they present
many features that hold in QCD: asymptotic freedom, 
gauge invariance, existence of 
a confining potential between non gauge invariant states 
(that should get eventually screened by the dynamical constituents),
and non-trivial topological structure (instantons, anomalies, $\theta$ vacua).
Moreover, a pleasant feature of these  models is the possibility
of performing a systematic $1/N$ expansion around the large-$N$
saddle point solution~\cite{D-D-L,Witten-2,C-R}.

In two-dimensional $CP^{N-1}$ models one defines 
the topological charge density operator 
\begin{equation}
q(x) = {i\over2\pi}\,\epsilon_{\mu\nu}\, \overline{D_\mu z} D_\nu
z ={1\over2\pi}\,\epsilon_{\mu\nu}\, \partial_\mu A_\nu,
\end{equation}
and its two-point correlation function $G_q(x-y) = \langle q(x) q(y) \rangle$.
$q(x)$ is a renormalization group invariant operator,
i.e. it does not have an anomalous dimension.
Like the topological charge density of QCD,
$q(x)$ transforms as
\begin{equation}
\Theta q(x)  = - q(\theta x) = - q(x_1,-x_2)
\end{equation}
under reflection symmetry. 
Therefore, as a consequence of reflection positivity,
$G_q(x)<0$ for $x\neq 0$.

Let us introduce the second-moment correlation length $\xi$ associated with
the two-point correlation function of the operator $P_{ij}(x) \equiv  \bar{z}_i(x) z_j(x)$,
\begin{equation}
G_P(x-y) = \langle {\rm Tr}\, P(x) P(0) \rangle, \label{GP}
\end{equation}
\begin{equation}
\xi^2 ={ \int d^2 x \;\case{1}{4} x^2 G_P(x)\over 
\int d^2 x  \;G_P(x) }.
\end{equation}
We will use $\xi$ as length scale in the 
following calculations \footnote{The second-moment correlation length of $G_P(x)$
turns out to be more suitable for a $1/N$-expansion than the mass-scale
determined from the large-distance exponential decay of $G_P(x)$, due to its
analytical properties in $1/N$~\cite{C-R}.}.

The Fourier transform of $G_q(x)$ can be written in terms of the propagator of the
composite field $A_\mu$, $P^A_{\mu\nu}(p)$,
\begin{equation}
\widetilde{G}_q(p) = {1\over 4\pi^2} \epsilon_{\mu\nu} \epsilon_{\rho\sigma}
p_\mu p_\rho P^A_{\nu\sigma}(p).
\end{equation}
Substituting the leading large-$N$ expression of $P^A_{\mu\nu}$~\cite{D-D-L,Witten-2,C-R},
one finds
\begin{eqnarray}
N \widetilde{G}_q(p) &=& {1\over 2\pi} p^2 \left[ u(p) \ln {u(p)+1\over u(p)-1} - 2\right]^{-1},
\label{tG}\\
u(p) &=& \sqrt{ 1 + {2\over 3 p^2\xi^2} }.\nonumber 
\end{eqnarray}
It is worth noting that in two-dimensional $CP^{N-1}$ models
the two-point correlation function of $q(x)$ is also related to the
correlation of two Wilson loops constructed 
with the abelian field $A_\mu$ in the limit of small area~\cite{C-R}.
 
Since it is convenient to work with dimensionless quantities, 
we define 
\begin{equation}
B(k) \equiv  \xi^2 N \widetilde{G}_q(p=k/\xi),
\label{bk}
\end{equation}
i.e. we use $\xi$ as unit of length.
$B(k)$ has the following asymptotic behaviors
\begin{eqnarray}
B(k)&=& {k^2\over 2\pi \ln (6 k^2/e^2)} + O\left( {1\over \ln k}\right),
\label{yinf}\\
B(k) &=& {1\over 2\pi} + {3\over 10\pi} k^2  - {27\over 350\pi} k^4 + O(k^6),
\label{y0}
\end{eqnarray}
for large and small momentum respectively.

The singular behavior of $G_q(x)$ at small distance is already apparent
from the asymptotic behavior (\ref{yinf}) of its Fourier transform. 
The calculation of the large-$N$ limit of 
$G_q(x)$ requires to perform the Fourier transform
of the expression (\ref{tG}).
As before let us introduce the dimensionless quantity:
\begin{equation}
C(x) \equiv \xi^4 G_q(x\xi) = 
\int {d^2 k\over (2\pi)^2} e^{ik\cdot x} B(k).
\label{cz}
\end{equation}
The moments of $C(x)$ 
\begin{equation}
\overline{\chi}_{q,j} \equiv \int d^2x\;(x^2)^j C(x) =
\xi^{2(1-j)} \chi_{q,j} 
\end{equation}
(where $\chi_{q,j} \equiv  \int d^2 x \,(x^2)^{j} G_q(x)$), 
and therefore of $G_q(x)$,
can be easily obtained  from 
the expansion of $B(k)$ in powers of $k^2$, cf. Eq.~(\ref{y0}).
In the large-$N$ limit one finds
\begin{eqnarray}
\xi^2 \chi_q &=& {1\over 2\pi N} + O\left( {1\over N^2}\right),\\
\chi_{q,1} &=& -{6\over 5\pi N } + O\left( {1\over N^2}\right),
\end{eqnarray}
etc...
We mention that $\chi_q$ and $\chi_{q,1}$ are known to 
$O(1/N^2)$~\cite{Luscher-chicpn,C-R-chicpn}.

By rotational invariance $C(x)$ depends only on $r\equiv |x|$:
\begin{equation}
C(x) = {1\over 2\pi} \int_0^\infty k dk  J_0(kr) B(k). 
\end{equation}
In order to evaluate $C(x)$ for finite $r\equiv |x|$, it is 
convenient to modify the integration contour in the complex $k$ plane,
moving it along the imaginary axis (when $r>0$)~\cite{C-R}.
For $r>0$ one can then write
\begin{equation}
C(x) = - {1\over 2\pi^2} \int_{\sqrt{2\over 3}}^\infty 
dt K_0(tr) t^3 v(t) \left[ \left( v(t)\ln{1+v(t)\over 1-v(t)} - 2\right)^2 + \pi^2 v(t)^2\right]^{-1},
\label{Kr}
\end{equation}
where $v(t) = \sqrt{ 1 - \case{2}{3t^2}}$. Since $K_0(x)>0$,
Eq.~(\ref{Kr}) shows that $C(x)<0$ for $r>0$ as expected. 
In Fig.~\ref{fig} we show the function $C(x)$.
The integral representation (\ref{Kr}) for $C(x)$ holds only for $r>0$.
For $r=0$ the contour rotation leading to the integral representation 
(\ref{Kr}) misses the contribution from the path at infinite distance, that is not
suppressed anymore and that should represent the positive contact term.

The integral representation (\ref{Kr}) allows us to derive the asymptotic behaviors of $C(x)$.
At large distance $C(x)$ decays exponentially
\footnote{We used the asymptotic behavior 
\begin{equation}
K_0(x)= \left({\pi\over  2x}\right)^{1/2} e^{-x}\left[ 1 + O\left( {1\over x}\right)\right].
\end{equation}
for $x\rightarrow \infty$.}:
\begin{equation}
C(x) =  -{1\over 24\pi} {e^{-\sqrt{2\over 3}r}\over r^2} \left[ 1 + O\left({1\over r}\right)\right].
\end{equation}
For $r\rightarrow 0$, $C(x)$ diverges as
\begin{equation}
C(x) = - {1\over 2\pi^2} \,{1\over r^4 \left(\ln r\right)^2 } 
\left[ 1 + O\left( {1\over \ln r}\right)\right].
\label{leadterm}
\end{equation}
One can infer this short-distance behavior also by 
calculating the leading order of perturbation theory 
%(derived using the Gaussian propagator $P_G(x)= \case{1}{\pi}\ln \left( \case{1}{|x|}\right)$)
that is given by 
\begin{equation}
-{g^2\over 2\pi^4 r^4}.
\end{equation} 
Then, using renormalization group
arguments, one replaces the coupling $g$ with a running coupling
constant $g(r)\approx \pi/\ln (\Lambda/r)$.

The diverging negative integral of $C(x)$ for $r>0$ must be compensated
by a diverging positive contribution of the contact term at $r=0$
(that got lost in the integral contour rotation performed to evaluate $C(r)$ for $r>0$)
so that 
\begin{equation}
\int d^2 x\; C(x) = B(0)={1\over 2\pi}.
\end{equation}
The integrals for $r>r_0>0$ are finite; for example
numerically one finds
$\int_{r>1} d^2x \;C(x) = -0.0503...$,
$\int_{r>1/2} d^2 x\; C(x) = -0.1653...$,
etc... 
Notice that as a consequence of the $r^{-4}$ short-distance behavior
and the positivity of $\chi_q$, we have formally
$\int_{|x|<\delta} d^2 x\;C(x) \longrightarrow + \infty$ 
for $\delta\rightarrow 0$. Thus a $\delta$-like distribution cannot
represent the contact term. The behavior at $x=0$ should be described by
more complicated distributions acting in a finite interval
\footnote{ An example of such distributions may be
\begin{equation}
{\rm lim}_{\varepsilon\rightarrow 0^+} 
\left[ P_\varepsilon(\partial) \delta(\vec{x}) - 
\lambda_\varepsilon {1\over |x|^4} f\left( \ln |x| \right)
\theta(x-\varepsilon)\right]
\end{equation}
where the polynomial $P_\varepsilon(\partial)$ and $\lambda_\varepsilon$ are  appropriate functions of
$\varepsilon$ and the limit $\varepsilon\rightarrow 0$
must be considered in a weak sense, i.e. after  performing the integral with the test function.}.
Similar considerations hold in QCD.

For comparison one may calculate the large-$N$ limit of $G_P(x)$, cf. Eq.(\ref{GP}).
Since the operator $P_{ij}(x)$ is not renormalization group invariant,
in order to construct a renormalization group 
invariant function, one may define
\begin{equation}
f(x)\equiv {G_P(x)\over \chi_P/\xi^2}
\end{equation}
where $\chi_P=\widetilde{G}_P(0)=
\case{6}{\pi} g^2\xi^2$ is the corresponding susceptibility,
and therefore
$\chi_P/\xi^2$ is  the renormalization constant at zero momentum of the operator $P_{ij}(x)$.
In the large-$N$ limit one finds (in units of $\xi$)
\begin{equation}
\widetilde{f}(k) = {1\over  3 k^2 u(k)} \ln{u(k)+1\over u(k)-1},
\end{equation}
whose asymptotic behaviors are
\begin{eqnarray}
\widetilde{f}(k)&=& {\ln(6k^2)\over 3k^2} + O\left( {\ln k \over k^4}\right),\\
\widetilde{f}(k) &=& 1 - k^2 +  O(k^4)\label{wfkl}
\end{eqnarray}
for large and small momentum respectively.
After some manipulations, for $r\equiv |x|>0$
\begin{equation}
f(x)
= {1\over 3\pi} \int_{\sqrt{2\over 3}}^\infty dt {K_0(tr)\over t v(t)}
\end{equation}
(where again an integral contour rotation of the Fourier transform of $\widetilde{f}(k)$ has been 
performed),
with the following asymptotic behaviors 
\begin{eqnarray}
&f(x) \sim {1\over r}e^{-\sqrt{2\over 3}r}\qquad\qquad
&{\rm for} \qquad r\rightarrow \infty,\\
& f(x) \sim  \left(\ln r\right)^2 
\qquad\qquad
&{\rm for} \qquad r\rightarrow 0  .
\end{eqnarray} 
In the integral contour rotation the contribution to the contact term 
of the path at infinite distance is again lost. 
Comparing with the behavior of the topological charge density correlation function,
we note that the singularity at $r=0$ is much softer in this case.
This is already apparent by comparing the large momentum behaviors of
$\widetilde{f}(k)$ and $B(k)$, cf. Eqs.~(\ref{wfkl}) and (\ref{yinf}).

\section{ The topological charge density correlation function on the lattice.}
\label{sec4}

In a pure gauge theory
$q(x)$ is a renormalization group invariant operator,
thus $G_q(x)$ is the 
universal continuum limit of corresponding lattice correlations
$\langle q_L(x) q_L(y) \rangle$,
when expressed in the appropriate 
units. In the presence of fermions a nontrivial anomalous 
dimension is generated
due to the mixing with the operator $\partial_\mu j_\mu^5(x)$~\cite{S-V}.
One may still define a universal RG invariant
function from the topological charge density correlation function, using
the fact that the anomaly equation (in the chiral limit)
$\partial_\mu j_\mu^5(x)=i2N_f q(x)$ does not get renormalized~\cite{A-D-P-V}.

Further subtleties are present on the lattice.
For example let us consider the case of a pure gauge theory.
A more careful analysis of the continuum limit of the matrix elements
of a generic lattice discretization 
$q_L(x)$ of $q(x)$,
such as Eq.~(\ref{Q^L}),
leads to the relation~\cite{C-D-P}
\begin{equation}
q_{L}(x)=a^4 Z(g) q(x)+O(a^{6}), 
\end{equation}
where $a$ is the lattice spacing, $g$ is the bare lattice coupling and
$Z(g)=1+O(g^2)$ a finite  renormalization function.
Thus we have
\begin{equation}
G^{L}_q(x) = a^8 Z(g)^2 G_q(ax)\left[ 1 + O(a^{2})\right]
\qquad {\rm for} \quad ax\neq 0.
\end{equation}
In order to evaluate the moments of $G_q(x)$ and in particular
the topological susceptibility, one also needs the contribution at
$x=0$.  This is strongly affected by lattice artifacts 
that eventually become dominant in the continuum limit.
So a careful subtraction is required (see e.g. Ref.~\cite{D-V,A-C-D-G-V}).
In order to overcome these problems, geometrical definitions 
have been proposed~\cite{B-L,L-gm}, that represent
appropriate interpolations among the discrete lattice variables,
and are in general non-analytical and non-single-valued functions of
the lattice variables.
These lattice estimators are not affected by renormalizations.
On the other hand the corresponding topological susceptibility, and probably also
higher moments, turns out to be  sensitive to unphysical lattice
defects~\cite{L-dl}, that may spoil their continuum
limit in some cases (see also Refs.~\cite{P-T,G-K-S-W-2,R-R-V}).

In order to investigate how $G_q(x)$ is recovered in the continuum limit of lattice
correlations $G_q^L(x-y)\equiv \langle q_L(x) q_L(y) \rangle$,
we consider a lattice formulation of the two-dimensional $CP^{N-1}$ models.
We consider the lattice action~\cite{R-S,D-H-M-N-P}:
\begin{equation}
S_L = -N\beta\sum_{n,\mu}\left( 
   \bar z_{n+\mu}z_n\lambda_{n,\mu} +
   \bar z_nz_{n+\mu}\bar\lambda_{n,\mu} - 2\right),
\label{basic}
\end{equation}
where $\beta=1/g$, $z_n$ is a $N$-component complex vector, constrained by
the condition $\bar z_nz_n = 1$,
and $\lambda_{n,\mu}$ is a ${\rm U}(1)$ variable.
One can easily prove that site- and link-reflection positivity holds for the 
lattice action (\ref{basic}).
The  lattice formulation (\ref{basic}) of $CP^{N-1}$ models 
turns out to be particularly convenient for a large-$N$ 
expansion~\cite{D-H-M-N-P,D-M-N-P-R,R-V,C-R-revcpn,R-R-V}.

In an infinite lattice (free boundary conditions are assumed)
one may consider the following discretization of  
the topological charge density operator 
\begin{equation}
q_L(n) = {1\over4\pi}\epsilon_{\mu\nu} (\theta_{n,\mu} + 
   \theta_{n+\mu,\nu} - \theta_{n+\nu,\mu} - \theta_{n,\nu}),
\label{qgeomtheta2}
\end{equation}
where $\theta_{n,\mu}$ is the phase of the field
$\lambda_{n,\mu}$, i.e.
$\lambda_{n,\mu} \equiv e^{i\theta_{n,\mu}}$.
Using the property of $q_L(n)$ under site-reflection one can prove that
$G_q^L(x)\equiv \langle q_L(x)q_L(0)\rangle $ must be
 negative for $|x|>0$ (at least along 
the directions of the lattice).
At large $N$ one can explicitly show that $q_L(n)$ has the correct
continuum limit, and  no lattice renormalizations 
are necessary~\cite{D-M-N-P-R,R-R-V}. 
Thus in the continuum limit  and for $|x|>0$ one expects
\begin{equation}
C^L(x)\equiv \xi^4 N G_q^L(x\xi) = C(x) + O(\xi^{-2}),
\label{clx}
\end{equation}
where $C(x)$ is the continuum function defined in Eq.~(\ref{cz}),
and $\xi$ the second-moment correlation length associated with the lattice
correlation function $G_P(x) \equiv \langle {\rm Tr}\,P(x) P(0)\rangle$.

$G_q^L(x)$ can be calculated in the large-$N$ limit.
Straightforward calculations lead to
\begin{equation}
N \widetilde{G}_q^L(k) = \sum_x e^{ik\cdot x} N G_q^L(x) = 
{1\over (2\pi)^2} \hat{k}^2 \Delta_{(\lambda)}(k),
\label{Gqlk}
\end{equation}
where~\cite{D-H-M-N-P,D-M-N-P-R,R-V}
\begin{equation}
\Delta_{(\lambda)}^{-1}(k) = \int_{-\pi}^{\pi} {d^2q\over (2\pi)^2} 
{2\sum_\mu \cos q_\mu\over \widehat{q}^2+m_0^2} 
-\int_{-\pi}^{\pi} {d^2q\over (2\pi)^2} {4\sum_\mu \sin^2\left(q_\mu + k_\mu/2\right)
\over \left[\widehat{q}^2+m_0^2\right] 
\left[\widehat{(q+k)}^2+m_0^2\right] },
\end{equation}
and $\widehat{q}^2\equiv \sum_\mu \widehat{q}^2_\mu \equiv 4\sum_\mu \sin^2(q_\mu/2)$.
The parameter $m_0^2$ is related to $\beta$ by 
\begin{equation}
\beta = \int_{-\pi}^{\pi} {d^2q\over (2\pi)^2} 
{1\over \widehat{q}^2+m_0^2 } =
{1\over 2\pi(1 + m_0^2/4)} K\left( {1\over 1 + m_0^2/4}\right)
\end{equation}
($K$ is the standard elliptic function),
and to the second-moment 
correlation length associated with $G_P(x)$ by 
\begin{eqnarray}
&&\xi^2 = {1\over 2\widetilde{G}^L_P(0)}{\partial^2 \over \partial k_1^2}
\widetilde{G}^L_P(k)|_{k=0}=
-\Delta_{(\alpha)}(0)  {1\over 2}
{\partial^2 \over \partial k_1^2} \Delta_{(\alpha)}^{-1}(k)|_{k=0},
\label{xilat}\\
&&\Delta_{(\alpha)}^{-1}(k) = \int_{-\pi}^{\pi} {d^2q\over (2\pi)^2} 
{1\over \left[\widehat{q}^2+m_0^2\right] 
\left[\widehat{(q+k)}^2+m_0^2\right] }.\nonumber 
\end{eqnarray}
$G_q^L(x)$ can be evaluated by Fourier transforming 
$\widetilde{G}_q^L(k)$. 
The lattice topological susceptibility is obtained by summing
$G_q^L(x)$ over the lattice sites,
\begin{equation}
\chi^L_q \equiv \sum_x G^L_q(x) = \widetilde{G}^L_q(0).
\end{equation}
An analysis of the above large-$N$ expressions leads to the 
following  main results.

\noindent 
(i) The continuum limit of $\xi^2 \widetilde{G}^L(k)$ at $k\xi$ fixed is
$B(k\xi)$, cf. Eq.~(\ref{bk}).  Indeed  at large $\xi$
\begin{equation}
\xi^2 G^L_q(k) = B(k\xi)+O(\xi^{-2}). 
\label{xigl}
\end{equation}
This may be seen by performing an asymptotic
expansion  of $\widetilde{G}^L_q(k)$ (at fixed $k\xi$) in powers of $\xi^{-2}$,
following the procedure outlined in Ref.~\cite{C-R-revcpn}.

\noindent 
(ii) $G_q^L(x)$ is  negative everywhere for $x\neq 0$,
consistently with reflection positivity.

\noindent 
(iii) At fixed physical distance $r=x/\xi >0$ the continuum limit 
exists and it is given by $C(r)$, in agreement with Eq.~(\ref{clx}).
Notice that the convergence is not uniform in $r$.
Moreover, from  Eq.~(\ref{xigl}) it follows that 
\begin{equation}
\xi^{2(1-j)} \chi^L_{q,j} = \overline{\chi}_{q,j} + O(\xi^{-2})
\end{equation}
where $\chi^L_{q,j}$ are the lattice moments of $G^L_q(x)$ and 
$\overline{\chi}_{q,j}$ are the moments of $C(r)$. 

\noindent 
(iv) $G_q^L(0)$ compensates the negative sum $\sum_{x\neq 0} G^L_q(x)$
and makes $\chi_q$ positive. Moreover, at large $\xi$, where $\xi\sim \exp (2\pi\beta)$,
\begin{equation}
G_q^L(0) \sim {1\over (\ln\xi )^2}\sim {1\over  \beta^2},
\end{equation}
giving rise to a positive contact term in the continuum limit.

In the Table we report some results for $G_q^L(0)$, and the scaling quantities
$C^L(x)$ (for some values of $x$) and $\xi^2\chi^L_q$.

Among the lattice techniques  used to  determine $\chi_q$ in Monte
Carlo simulations, there are also non-local estimators (see e.g. cooling).
Since $\chi_q = {1\over V} \langle Q^2 \rangle$,
where $Q$ is the total topological charge ($Q= \sum_x q_x$)
and $Q$ is stable (approximately on the lattice) 
under local changes, in the cooling procedures one changes appropriately the configuration
to read $Q$ and therefore to determine $\chi_q$. 
But of course this procedure does not leave the
two-point correlation function invariant: only its zero-mode should be left intact.

In conclusion we have seen that the continuum limit of the correlation function
$G_q^L(x)$ is regular at fixed physical distance and it is given by
$G_q(x)$. 
Also its moments $\chi_{q,j}$
have a regular continuum limit. On the other hand,
a singular behavior is found at $x=0$ consistently with reflection positivity and
positivity of $\chi_q$. 
These features should also characterize the Euclidean correlation  function of the
topological charge density in the continuum limit of lattice QCD.

\acknowledgments

I gratefully acknowledge useful 
discussions with Mihail Mintchev,
Haralambos Panagopoulos, Andrea Pelissetto and Paolo Rossi.

% ========================= TABLES =========================

\begin{table}
%\squeezetable
\caption{
The large-$N$ limit of $N G_q^L(0)$, 
$C^L(x,0)\equiv \xi^4 N G_q^L(x\xi,0)$ and $\xi^2 N \chi_q^L$ 
for $\xi=1,2,4,8,16,\infty$
(corresponding to $\beta=0.41829..., 0.52868...,0.63901...,0.74933...,0.85965...,
\infty$ 
respectively). 
For $\xi=\infty$ the continuum results are recovered.
}
\label{}
\begin{tabular}{cr@{}lr@{}lr@{}lr@{}lr@{}lr@{}l} 
\multicolumn{1}{c}{$\xi$}&
\multicolumn{2}{c}{$N G_q^L(0)$}&
\multicolumn{2}{c}{$C^L(1/4,0)$}&
\multicolumn{2}{c}{$C^L(1/2,0)$}&
\multicolumn{2}{c}{$C^L(1,0)$}&
\multicolumn{2}{c}{$C^L(2,0)$}&
\multicolumn{2}{c}{$\xi^2 N \chi_q^L$}\\
\tableline \hline
1 & 0&.5371 &  &&  &&  $-$0&.07888 & $-$0&.001376 & 0&.199791 \\
2 & 0&.2751 &  &&  $-$0&.8148 & $-$0&.01634 & $-$0&.001257 & 0&.171593 \\
4 & 0&.1759 &  $-$9&.314 &  $-$0&.1713 & $-$0&.01615 & $-$0&.001160 & 0&.162993 \\
8 & 0&.1277 &  $-$1&.790 &  $-$0&.1728 & $-$0&.01547 & $-$0&.001136 & 0&.160316 \\
16 & 0&.09995 &  $-$1&.808 &  $-$0&.1662 & $-$0&.01535 & $-$0&.001131 & 0&.159498 \\
$\infty$ & & &  $-$1&.718 &  $-$0&.1651 & $-$0&.01532 & $-$0&.001129 & 0&.159155 \\
\end{tabular}
\end{table}

\begin{figure}
\caption{
Plot of $\ln \left[ - C(x)\right]$ versus $r\equiv |x|$.
}
\label{fig}
\end{figure}

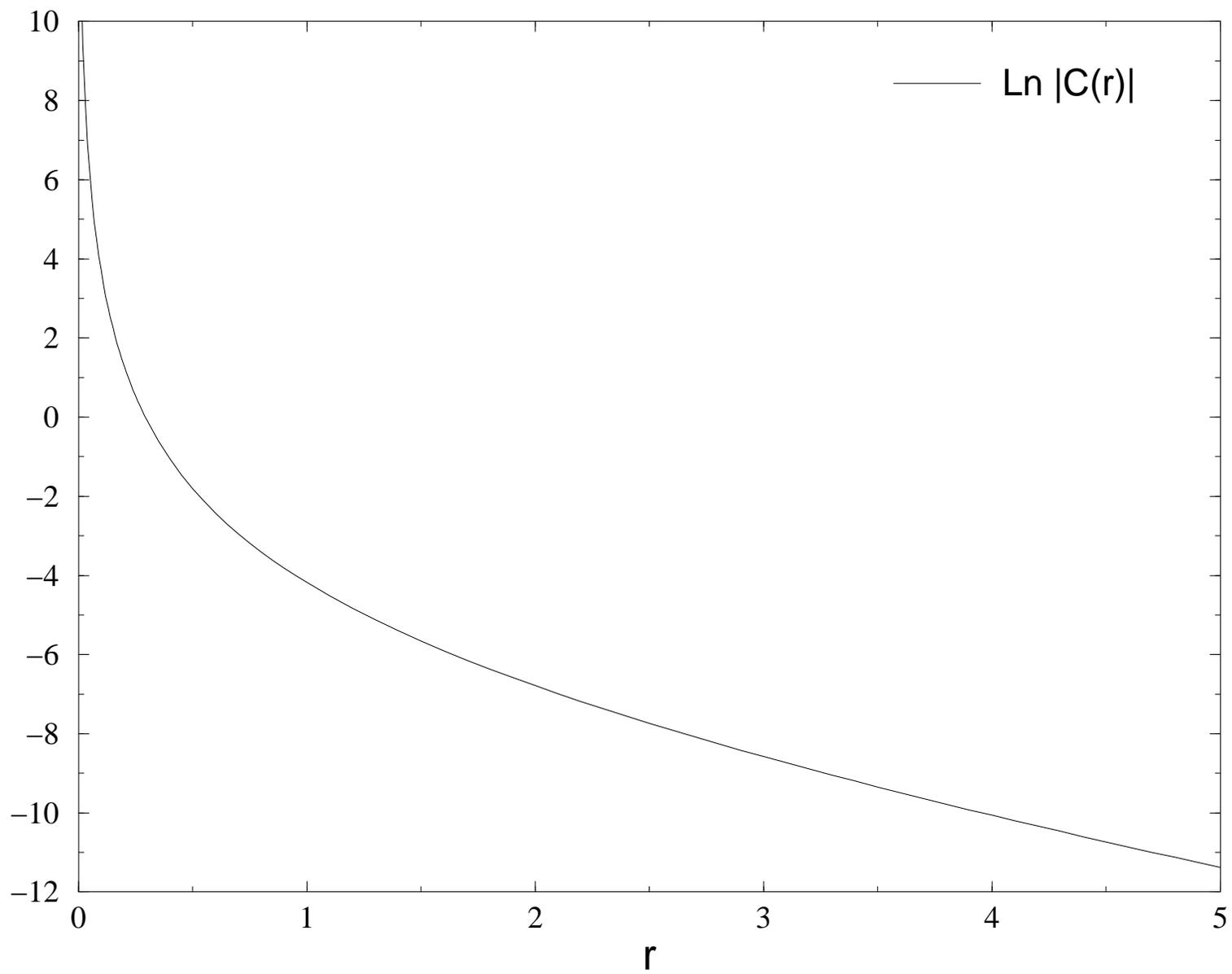
\end{document}